\documentclass[aps,prb,floats,twocolumn,10pt]{revtex4-1}
\usepackage{graphicx}
\usepackage{amssymb}
\usepackage{amsmath}
\usepackage{mathrsfs}
\newcommand\numberthis{\addtocounter{equation}{1}\tag{\theequation}}
\usepackage{color}
\usepackage[dvipsnames]{xcolor}
\usepackage{slantsc}
\usepackage{float}
\graphicspath{{./FIG_FINALI/}}
\usepackage{bbold}
\usepackage[caption=false]{subfig}
\usepackage{array}
\usepackage{hyperref}
\hypersetup{backref,pdfpagemode=FullScreen,colorlinks=true,allcolors=blue}
\usepackage{braket}
\usepackage{scrextend}
\usepackage[shortlabels]{enumitem}
\begin{document}

\title{CleaRIXS: A Fast and Accurate First-Principles Method for Simulation and Analysis of Resonant Inelastic X-ray Scattering}
\author{Subhayan Roychoudhury}
\email{roychos@tcd.ie}
\affiliation{The Molecular Foundry, Lawrence Berkeley National Laboratory, Berkeley CA 94720, USA}
\author{David Prendergast}
\email{dgprendergast@lbl.gov}
\affiliation{The Molecular Foundry, Lawrence Berkeley National Laboratory, Berkeley CA 94720, USA}

\begin{abstract}
Resonant Inelastic X-ray Scattering (RIXS), which probes the occupied and unoccupied electronic subspaces in an interrelated fashion, is one of the most detailed, complex and information-rich experimental techniques employed in the investigation of electronic structure across physics, chemistry and materials science. We introduce an ab-initio, accurate and efficient computational framework for simulation and analysis of RIXS spectra, which combines two diverse and established approaches to modeling electronic excited states. The Core-hole LinEAr-response RIXS (CleaRIXS) method not only ensures accurate incorporation of the interaction of electrons with core and valence holes, but also automatically maps the salient RIXS features to the relevant electronic excitations and de-excitations. Through direct comparison with previous methanol C K-edge RIXS measurements [J. Phys. Chem. A 120, 2260
(2016)], we show the efficacy of the formalism in modeling different regions of the RIXS spectrum and in gaining physical insight regarding their origins. The importance of including the valence electron-hole interactions is explored, in addition to the connection between CleaRIXS and determinant-based approaches for simulating X-ray absorption and non-resonant X-ray emission. CleaRIXS provides a robust and extendable framework for prediction and interpretation of RIXS processes and for the simulation of complex electronic excited states in general.
\end{abstract}

\maketitle

\section{Introduction}
The ability to decipher, analyze and manipulate electronic structure of materials has been one of the most promising achievements of modern human civilization. Besides their innate importance in terms of scientific enquiry, such efforts have immense technological potential in various practical aspects including electronics~\cite{MESSMER1985285, doi:10.1146/annurev-matsci-071312-121630, doi:10.1098/rsta.2013.0270}, drug discovery~\cite{Schaduangrat2020, Ye2022,Arodola2017}, discovery of magnetic materials~\cite{doi:10.1126/sciadv.1602241}, energy storage~\cite{doi:10.1126/sciadv.abm2422, https://doi.org/10.1002/eem2.12056}, renewable energy generation~\cite{doi:https://doi.org/10.1002/9783527813636.ch19, PhysRevLett.112.196401} to name a few. Consequently, over the years, various theoretical and experimental techniques have been developed for studying the electronic structure and properties of materials under various conditions. 
Electronic structure is typically investigated in terms of electronic excitations which can be of two kinds :
(1) charged excitation - involving addition or removal of an electron, and (2) neutral excitation - which conserves the total number of electrons. This latter process is complicated (within the quasiparticle picture) by electon-hole interactions. 

However, experimental investigation of electronic structure is not a straightforward process. For example, attempts to probe the valence orbitals directly by creating charged excitations, i.e., by simply removing (photoemission) or adding (inverse photoemission) an electron has its drawbacks. Due to the relatively low energy and delocalized nature of such excitations, the resolution of such processes is often marred by various additional intrinsic or induced excitations, due to strong electron-electron and electron-phonon scattering. Similar issues are encountered in low-energy optical exploration of neutral excitations. One alternative, which provides the additional advantage of species-specificity, is to excite the strongly bound core electrons using X-ray photons, revealing details of occupied electronic structure using X-ray emission spectroscopy (XES)~\cite{doi:10.1063/1.4905603,PhysRevB.97.125139,10.1093/jmicro/dfx132,D1CC03023J,10.1093/jmicro/dfy024,10.1117/12.616260} and unoccupied electronic structure using X-ray absorption spectroscopy (XAS)~\cite{doi:10.1021/cr9900681,doi:10.1021/acs.chemrev.7b00213,PhysRevLett.96.215502,Li_2016,doi:10.1021/acsami.1c11970,FREIWALD2004413,liu_weng_2016,C4CP05316H,C6CP02412B}. With the advent of synchrotron X-ray light sources, X-ray spectroscopy has gained widespread popularity in the past decades. However, an even more powerful technique that can probe charged and neutral excitations in an interdependent manner and provide fingerprints of electronic/chemical processes that are unavailable in standard XAS/XES is Resonant Inelastic X-ray Scattering (RIXS)~\cite{RevModPhys.83.705,RevModPhys.73.203,D0DT01782E,doi:10.1021/acs.jpclett.8b02757,YANG2018188,doi:10.1021/acs.chemmater.0c03930,https://doi.org/10.1002/eem2.12172,Yan2854}. For example, in transition metal oxide electrodes, RIXS has identified previously unknown and technologically important oxygen-redox states~\cite{doi:10.1021/acs.jpclett.8b02757}, which, in the conventional XAS spectrum, are hidden under metal-oxide hybridization features~\cite{https://doi.org/10.1002/eem2.12119}. Owing to this high information-content, and increased access to the elemental edges of low-Z elements, such as O or C, at soft X-ray light sources~\cite{doi:10.1063/1.4977592}, RIXS is gaining increasing popularity in probing the electronic properties of molecules, bulk materials and surfaces. 

The RIXS process amounts to exciting a core-electron using incident photons of varying frequencies and recording, as a function of frequency, the intensity of the emission resulting from the de-excitation of this intermediate core-excited state. Roughly speaking, from the electronic perspective, the incident frequency can be tuned to excite a core-electron of a given atomic element to some unoccupied orbital, probing the so-called conduction or virtual levels. Subsequently, the coherent de-excitation of such core-excited states proceeds by annihilation of the so-called core hole by an electron from an occupied orbital. In soft X-ray RIXS, this is a valence electron and the resulting final state comprises a valence-only electron-hole pair, despite the initial involvement of a core electron. If we ignore any other loss processes, the energy of this electron-hole pair matches the energy loss defined by the incoming and outgoing photon energies. More generally, we can think of the RIXS process, by analogy with Raman processes, as providing access to a spectrum of lower-energy excitations (electron-hole pairs~\cite{Jia_2019,SHIRLEY2000305}, vibrations~\cite{doi:10.1063/1.5092174,doi:10.1021/jp505628y}, charge-transfer excitations~\cite{PhysRevLett.96.067402,doi:10.1063/1.3367958}, magnons~\cite{PhysRevB.97.155144}, etc. and their various combinations) defined by the associated energy loss. In this work, we will focus mostly on the utility of RIXS to probe the valence electron-hole interactions. Since the RIXS process does not create the final state by exciting the ground state directly, certain valence-excited states which are inaccessible under direct optical excitation, may be visible using RIXS. Therefore, RIXS provides an invaluable validation for theoretical approaches designed to accurately model valence excitonic states. 

With the advent of new and advanced light sources and the refinement of detection-techniques, RIXS is becoming more and more popular and its usage is only expected to increase with time. Evidently, extraction of valuable electronic-structure information from such a complicated experimental process requires complementary theoretical research.
Techniques based on exact-diagonalization~\cite{PhysRevLett.105.177401,Tsutsui_2012,PhysRevB.85.064423} as well as quantum chemical methods~\cite{doi:10.1021/jp505628y,doi:10.1063/5.0010295,C9CP03688A} such as the configuration-interaction equation-of-motion coupled-cluster formalism have been shown to provide accurate description of the RIXS process, albeit at high computational expense. Additionally, density functional theory (DFT) based simulations using various levels of independent-particle approximation with\cite{Jia_2019,doi:10.1063/1.4977178} or without the core-hole effect have been formulated. However, such methods typically ignore the valence excitonic effects in the final state. To mitigate this shortcoming, electron-hole interactions have been incorporated in RIXS simulations by combining linear-response solutions for the intermediate core-excited state and the final valence-excited state\cite{PhysRevResearch.2.042003,PhysRevLett.80.794,PhysRevB.94.035163,PhysRevB.96.205116,doi:10.1021/acs.jctc.8b00211}, leveraging solutions to the Bethe-Salpeter equation (BSE)~\cite{PhysRevB.62.4927,PhysRevLett.80.794} or linear-response time-dependent DFT (LR-TDDFT)~\cite{doi:10.1142/9789812830586_0005}. However this can prove to be significantly expensive, especially since it involves all-electron (core and valence) calculations to provide explicit treatment of the core orbital. Additionally, use of linear response-based methods within the adiabatic approximation can incur inaccuracies in the oscillator strength~\cite{roychoudhury2021changes} if the excitation significantly alters the electron density (and consequently, the polarization). While typically this is not an issue with the low-lying valence excitations, due to the localized nature of the core-hole, the adiabatic approximation can prove to be inadequate for core-excitations.  

Evidently, an accurate and efficient ab-initio computational technique for modeling RIXS must be rooted in methods capable of simulating electronic excitations and de-excitations.
For this, the two most widely practised approaches built on top of the Kohn-Sham (KS) density-functional theory (DFT) framework are
\begin{enumerate}[(a)]
    \item the excited-state specific constrained-occupation approaches which directly approximate each excited-state as a Slater-determinant (SD) by imposing non-aufbau occupation within various possible levels of self-consistency~\cite{doi:10.1021/jp801738f,doi:10.1063/1.5018615,Hait2021,Roychoudhury2020,PhysRevLett.96.215502}, and
    \item the response based perturbative methods which operate by creating explicit e-h pairs in the ground-state SD~\cite{doi:10.1142/9789812830586_0005,BAUERNSCHMITT1996454,PhysRevLett.80.794,PhysRevB.62.4927}.
\end{enumerate}
It must be noted that an efficacious theoretical contender for RIXS must satisfy two distinct criteria simultaneously: 
\begin{enumerate}
    \item it must provide accurate spectral plots at acceptable computational expense by efficiently modeling the attraction between the electrons and core-hole (in the intermediate state) and the valence hole (in the final state), and
    \item it must help analyze the spectrum by associating the relevant electronic excitations/de-excitations with the spectral features.
\end{enumerate}

In this paper, we present Core-hole LinEAr-response RIXS (CleaRIXS), an efficient theoretical formalism with relatively lower computational demand which readily fulfills the aforementioned criteria by treating the electron-hole (e-h) interaction in the intermediate and in the final state with the help of two distinct mechanisms. In particular, the interaction in the intermediate core-excited state is modeled with the help of a constrained-occupation based single-determinant approach (belonging to category (a) mentioned above) that uses a modified pseudopotential to simulate the effects of the core-hole while that in the final valence-excited state is modeled by employing the perturbative linear-response treatment (belonging to category (b) mentioned above) applied on the ground state.

For the treatment of the final state, this method can, in principle, work in conjunction with any response-based technique or software~\cite{MARINI20091392,Sangalli_2019,DESLIPPE20121269, doi:10.1063/5.0055522} capable of modeling valence electronic excitations. Therefore, thanks to the ever-increasing availability of high-quality experimental RIXS data, CleaRIXS can complement the active research on the development of various advanced perturbative approximations (e.g. formulation of exchange-correlation kernels that include excitonic effects~\cite{PhysRevResearch.2.013091,Byun_2020,PhysRevLett.88.066404,PhysRevB.76.161103}) by providing a potent avenue for testing and validating their accuracy.
Representation of the intermediate state entirely within the KS framework ensures computational efficiency while a linear-response approach for the final state warrants accurate incorporation of the valence e-h interaction and possible excitonic effects, thereby satifying criterion (1). 
On the other hand, as shown in this paper, such a division-of-representation automatically satisfies criterion (2) by mapping the RIXS features to the relevant electronic excitations and de-excitations.

\section{Method}
For an incident (emitted) photon with energy $\hbar \omega_{\rm{in}}$ ($\hbar \omega_{\rm{out}}$) and polarization along the unit vector $\mathbf{\epsilon}_p$ ($\mathbf{\epsilon}_q$), denoting the ground state (GS) by $\ket{\mathrm{GS}}$ , the set of core-excited states by $\{\ket{X_k}\}$, and the total energy of any state $\ket{S}$ by $E_S$, the Kramers-Heisenberg formula~\cite{sakurai2006advanced, PhysRevB.49.5799} can be used to approximate the RIXS cross-section~\footnote{Only the resonant term has been included here. The normal inelastic scattering, which is a first-order process in electron-photon interaction, and the non-resonant anomalous scattering, which does not contain the input-photon frequency in the denominator, have been ignored. See ref.~\cite{PhysRevB.49.5799} for details.} as
\begin{align*}\label{eq1}
     &{} \sigma_{p,q}(\omega_{\rm{in}} , \omega_{\rm{out}}) = \sum_f \left\rvert T_{\omega_{\rm{in}},F}^{p,q}\right\rvert^2  \delta(\omega_{\rm{in}}+E_{\rm{GS}}-\omega_{\rm{out}}-E_F) \numberthis,
\end{align*}
with 
\begin{align}\label{eq1a}
   T_{\omega_{\rm{in}},F}^{p,q} =  \sum_k \frac{\braket{F|\hat{O}^\dagger_p|X_k}\braket{X_k|\hat{O}_q|\rm{GS}}}{\omega_{\rm{in}}-(E_k - E_{\rm{GS}})+i\Gamma_k}
\end{align}
where $\hbar$ has been set to unity. Here $O_q=\mathbf{\epsilon}_q \cdot \hat{\mathbf{R}}$ ($\hat{\mathbf{R}}$ being the many-body position operator) is the dipole transition operator and the imaginary term $i\Gamma_k$ takes care of the spectral broadening caused by the core-hole lifetime and instrument resolution (which we approximate as independent of $k$ here). The set of final states $\{\ket{F}\}$ encompasses the GS as well as all the valence-excited states, including those with an unbound electron. The $F=\rm{GS}$ term, in fact, represents the elastic contribution where, following the core-excitation, the system reverts back to the ground state. On the other hand, when $\ket{F}$ contains an unbound electron, we obtain the non-resonant XES cross-section, which is discussed in detail below.

The GS of a system with $N$ valence electrons and one (relevant) core electron can be written, by reference to a Hilbert space of single-particle orbitals (the solutions of the Kohn-Sham equations consistent with the GS electron density), within the single configuration approximation as a Slater determinant,
\begin{align}
    \ket{\rm{GS}}=\left(\prod_{i=1}^N a_i^\dagger\right) a_c^\dagger\ket{0},
\end{align}
where $a_c^\dagger$ creates an electron in the relevant core orbital $\ket{c}$ while $a_i^\dagger$ creates as electron in the $i$-th valence GS orbital $\ket{i}$. On the other hand, in analogy with the configuration-interaction singles method, each valence-excited final state (i.e., any final state other than $\ket{F}=\ket{\rm{GS}}$) can be written as a linear combination of Slater determinants obtained by creating an electron in an unoccupied orbital (say, $\alpha$) and a hole in an occupied orbital (say, $\beta$) of the GS Hilbert space. Thus any such final state is expressed as 

\begin{align}\label{f_expansion}
    \ket{F} = \sum_{\alpha,\beta} C^F_{\alpha , \beta} a^\dagger_\alpha a_\beta  \left(\prod_{j=1}^N a_j^\dagger\right)a_c^\dagger\ket{0},
\end{align}
where $C^F_{\alpha,\beta}$ is the coefficient of linear expansion.

Now, within a frozen orbital approximation for the core-excited state, for any $k>N$, an intermediate state can be written as 
\begin{align}
    \ket{X_k} =  \tilde{a}_k^{\dagger} \prod_{i=1}^N \tilde{a}_i^{\dagger}\ket{0},
\end{align}
where $\tilde{a}_i^\dagger$ denotes the creation operator for $\ket{\tilde{i}}$, the $i$-th valence orbital corresponding to the self-consistent field obtained in absence of the relevant core electron -- so-called core-hole excited-state orbitals. Here we have ignored those intermediate states, which, in addition to the core-hole, contain holes in the valence subspace (this has been referred to as the $f^{(1)}$ approximation in our previous work)~\cite{PhysRevLett.118.096402,PhysRevB.97.205127}. 

Finally, the many-electron transition operator $\hat{O}$ is written as
\begin{align}\label{eq5}
    \hat{O}_q = \sum_{m,n} \braket{m|\hat{o}_q|n}a_m^\dagger a_n, 
\end{align}
where $\hat{o}_q=\mathbf{\epsilon}_q \cdot \hat{\mathbf{r}}$ ($\hat{\mathbf{r}}$ being the single-particle position operator) is the single-particle dipole transition operator. Omitting the $q$-dependence for notational simplicity, for excitation of a core-electron, Eq.~\ref{eq5} can be re-written as 
\begin{align}
    \hat{O}=\sum_m \braket{m|\hat{o}|c}a^\dagger_ma_c=\sum_m o_m a^\dagger_m a_c,
\end{align}
where $o_m=\braket{m|\hat{o}|c}$. 

Significant computational simplification of Eq.~\ref{eq1} can be achieved with the help of linear algebra. Referring to the term $\braket{X_k|\hat{O}|\rm{GS}}$ as the absorption-amplitude $A^{a}_{\rm{GS}\rightarrow X_k}$, it can be shown~\cite{PhysRevB.100.075121} that

\begin{align}\label{eq7}
    \left(A^{a}_{\rm{GS}\rightarrow X_k}\right)^* &{}=\rm{det} \begin{bmatrix}
    \xi_{1,1} & \xi_{1,2} & \hdots & \xi_{1,N} & \sum\limits_{m>N}o^*_m \xi_{1,m}\\
    \vdots & \vdots & \ddots & \vdots & \vdots\\
   \xi_{N,1} & \xi_{N,2} & \hdots & \xi_{N,N} & \sum\limits_{m>N}o^*_m \xi_{N,m}\\
    \xi_{k,1} & \xi_{k,2} & \hdots & \xi_{k,N} & \sum\limits_{m>N}o^*_m \xi_{k,m}\\
    \end{bmatrix}, 
\end{align}
where $\xi_{i,j}=\braket{j|\tilde{i}}$ is the single-particle overlap term, between the ground and core-excited orbital spaces. For all $k>N$, Eq.~\ref{eq7} is solved with the help of a reference matrix using a relatively inexpensive algebraic trick shown in Ref.~\cite{PhysRevB.100.075121}.

Once again, omitting the dependence on the polarization ($p$) for simplicity, the emission amplitude $A^{e}_{X_k\rightarrow F}=\braket{F|\hat{O}^\dagger|X_k} =\braket{X_k|\hat{O}|F}^*$ can be written as
\begin{align}
    &{}A^{e}_{X_k\rightarrow F} = \sum_{\alpha,\beta}  \left(C^F_{\alpha,\beta}\right)^* Y^k_{\alpha,\beta},
\end{align}
where
\begin{widetext}
\begin{align*}
   Y_{\alpha,\beta}^k  &{}= \rm{det} \begin{bmatrix}
{\xi_{1,1}} & \hdots & {\xi_{1,\beta-1}} & {\xi_{1,\alpha}} &{\xi_{1,\beta+1}} &  \hdots & {\xi_{1,N}}  &  {\sum\limits_{m}o^*_m\xi_{1,m}}\\
\vdots & \ddots & \vdots & \vdots & \vdots & \ddots & \vdots & \vdots\\
{\xi_{N,1}} & \hdots & {\xi_{N,\beta-1}} & {\xi_{N,\alpha}} &{\xi_{N,\beta+1}} &  \hdots & {\xi_{N,N}}  &  {\sum\limits_{m}o^*_m\xi_{N,m}}\\
{\xi_{k,1}} & \hdots & {\xi_{k,\beta-1}} & {\xi_{k,\alpha}} &{\xi_{k,\beta+1}} &  \hdots & {\xi_{k,N}}  &  {\sum\limits_{m}o^*_m\xi_{k,m}}\end{bmatrix}\\
&{} = \rm{det} \begin{bmatrix}
{\xi_{1,1}} & \hdots & {\xi_{N,1}} & {\xi_{k,1}} \\
\vdots & \ddots & \vdots & \vdots\\
{\xi_{1,\beta-1}} & \hdots & {\xi_{N,\beta-1}} & {\xi_{k,\beta-1}} \\
{\xi_{1,\alpha}} & \hdots & {\xi_{N,\alpha}} & {\xi_{k,\alpha}} \\
{\xi_{1,\beta+1}} & \hdots & {\xi_{N,\beta+1}} & {\xi_{k,\beta+1}} \\
\vdots & \ddots & \vdots & \vdots\\
{\xi_{1,N}} & \hdots & {\xi_{N,N}} & {\xi_{k,N}} \\
\sum\limits_{\substack{m>N\\m\neq \alpha}}o^*_m \xi_{1,m}+o^*_\beta\xi_{1,\beta} & \hdots & \sum\limits_{\substack{m>N\\m\neq \alpha}}o^*_m \xi_{N,m} + o^*_\beta\xi_{N,\beta} & \sum\limits_{\substack{m>N\\m\neq \alpha}}o^*_m \xi_{k,m} + o^*_\beta\xi_{k,\beta} \\
\end{bmatrix},\\
\end{align*}
\end{widetext}

where, in the second step, we have simply transposed the matrix (since that preserves the determinant) and partitioned the each sum over $m$ into a term for the valence hole, $\beta$, and a sum over all the available conduction orbitals ($m>N$ satisfying $m\neq\alpha$).
By defining the row vectors of $Y_{\alpha,\beta}^k$ as
\begin{align}
    b^k_l = \begin{bmatrix}
    \xi_{1,l} & \hdots & \xi_{N,l} & \xi_{k,l},
    \end{bmatrix}
\end{align}

the expression can be simplified as

\begin{align*}\label{eq12}
   &{} Y_{\alpha,\beta}^k\\
&{}=b_1^k \wedge \hdots \wedge b_{\beta-1}^k \wedge b_\alpha^k \wedge b_{\beta+1}^k \wedge \hdots \wedge b_N^k \wedge\\
&{}\textrm{  } \left[\sum_{m>N} o^*_m b_m^k - o^*_\alpha b_\alpha^k + o^*_\beta b_\beta^k \right]\\
&{}=b_1^k \wedge \hdots \wedge b_{\beta-1}^k \wedge b_\alpha^k \wedge b_{\beta+1}^k \wedge \hdots \wedge b_N^k \wedge \left(\sum_{m>N}o^*_m b_m^k \right) \\
&{} + o^*_\beta \left[b_1^k \wedge \hdots \wedge b_{\beta-1}^k \wedge b_\alpha^k \wedge b_{\beta+1}^k \wedge \hdots \wedge b_N^k \wedge b_\beta^k \right]\\
&{}=b_1^k \wedge \hdots \wedge b_{\beta-1}^k \wedge b_\alpha^k \wedge b_{\beta+1}^k \wedge \hdots \wedge b_N^k \wedge \left(\sum_{m>N}o^*_m b_m^k \right) \\
&{} - o^*_\beta [b_1^k \wedge \hdots \wedge b_N^k \wedge b_\alpha^k]\\
&{} = D_1^{k,\alpha,\beta} - o^*_\beta D_2^{k,\alpha},\numberthis
\end{align*}
where $\wedge$ denotes a wedge-product (which is antisymmetric) and we have introduced the terms
\begin{align*}\label{eq13}
    &{}D_1^{k,\alpha,\beta}\\
&{}=b_1^k \wedge \hdots \wedge b_{\beta-1}^k \wedge b_\alpha^k \wedge b_{\beta+1}^k \wedge \hdots \wedge b_N^k \wedge \left(\sum_{m>N}o^*_m b_m^k \right)\numberthis
\end{align*}
and
\begin{align}\label{eqD2}
    D_2^{k,\alpha}=b_1^k \wedge \hdots \wedge b_N^k \wedge b_\alpha^k.
\end{align}
Note that in the penultimate step of Eq.~\ref{eq12}, we have used the fact that exchanging two rows changes the sign of a determinant. Thus, the task evaluating $Y^f_{\alpha,\beta}$ boils down to that of finding $D_1^{k,\alpha,\beta}$ and $D_2^{k,\alpha}$. 

The following steps exploit the fact that these determinants are defined only for linear vector spaces of finite dimension, i.e., the dimension of the associated square matrix, which permits any unknown row-vector to be expressed as a finite linear expansion in some reference basis. In order to find $D_1^{k,\alpha,\beta}$, we note that, for any $\alpha > N$, introducing a set of yet-unknown coefficients $\{\gamma\}$, we can write
\begin{align}\label{eq15}
    b_\alpha^k = \sum_{j=1}^{N} \gamma^k_{\alpha,j} b^k_j + \gamma^k_{\alpha,x} b^k_x 
\end{align}
where
\begin{align}
    b_x^k = \sum_{m>N} o^*_m b_m^k.
\end{align}
Now, plugging Eq.~\ref{eq15} into ~\ref{eq13} and noting that the determinant vanishes if any two rows are equal, we can find $D_1^{k,\alpha,\beta}$ as
\begin{align}\label{eq17}
    D_1^{k,\alpha,\beta} = \gamma^k_{\alpha , \beta} B_{\rm{ref}}^k,
\end{align}
where
\begin{align*}\label{eq18}
    B_{\rm{ref}}^k &{} = b_1^k \wedge \hdots \wedge b_N^k \wedge b_x^k = \left(A^a_{\rm{GS}\rightarrow X_k}\right)^*. \numberthis
\end{align*}
In Eq.~\ref{eq18}, we have used the fact that the determinant remains unchanged upon transposing a matrix.

Now, If there are $M$ number of KS orbitals in total, then, in order to find the unknown coefficients $\{\gamma\}$, we note that the $M-N$ linear equations Eq.~\ref{eq15} (one for each $\alpha>N$) can be written in matrix form as 
\begin{align}\label{eq19}
    \begin{bmatrix}
    b^k_{N+1}\\
    \vdots\\
    b^k_M\\
    \end{bmatrix} &{} = \begin{bmatrix}
    \gamma^k_{N+1,1} & \hdots & \gamma^k_{N+1,N} & \gamma^k_{N+1,x}\\
    \vdots & \ddots & \vdots & \vdots\\
    \gamma^k_{M,1} & \hdots & \gamma^k_{M,N} & \gamma^k_{M,x}\\ 
    \end{bmatrix} \begin{bmatrix}
    b^k_1\\
    \vdots\\
    b^k_N\\
    b^k_x\\
    \end{bmatrix}.
\end{align}
Denoting the three matrices (from left to right) in Eq.~\ref{eq19} as $\mathbf{B_1^k}$ , $\boldsymbol\gamma$ and $\mathbf{B_{\rm{ref}}^k}$, respectively, we can see that the unknown coefficients can be found by solving for $\boldsymbol\gamma$ using
\begin{align}\label{eq_for_gamma}
    \boldsymbol\gamma=\mathbf{B_1^k}\left(\mathbf{B_{\rm{ref}}^k}\right)^{-1}.
\end{align}

Evaluating $D_2^{k,\alpha}$ is relatively simpler. For any $\alpha > N$, we express $b_\alpha^k$ as a finite linear expansion
\begin{align}\label{eq19a}
    b_\alpha^k = \sum_{j=1}^{N+1} \kappa_{\alpha,j}^k b_j^k.
\end{align}
Plugging Eq.~\ref{eq19a} into Eq.~\ref{eqD2}, we find the simplified relation
\begin{align}\label{Eq.21}
    D_2^{k,\alpha,\beta} = \kappa_{\alpha,N+1}^k C^k_{\rm{ref}},
\end{align}
where $C^k_{\rm{ref}}$ is the determinant of  
\begin{align}
    \mathbf{C}^k_{\rm{ref}}=\begin{bmatrix}
    b^k_1\\
    \vdots\\
    b^k_{N+1}\\
    \end{bmatrix}.
\end{align}
The unknown coefficients $\{\kappa_{\alpha,j}^k\}$ can now be found by multiplying both sides of the equation (which is essentially the matrix form of Eq.~\ref{eq19a})
\begin{align}\label{Eq.23}
    \begin{bmatrix}
    b^k_{N+1}\\
    \vdots\\
    b^k_M\\
    \end{bmatrix} = \begin{bmatrix}
    0 & 0 & \hdots & 1\\
    \kappa^k_{N+2,1} & \kappa_{N+2,2} & \hdots & \kappa^k_{N+2,N+1}\\
    \vdots & \vdots & \ddots & \vdots\\
    \kappa^k_{M,1} & \kappa^k_{M,2} & \hdots & \kappa^k_{M,N+1}\\
    \end{bmatrix} \begin{bmatrix}
    b^k_1\\
    \vdots\\
    b^k_{N+1}\\
    \end{bmatrix}.
\end{align}
by $\left(\mathbf{C}^k_{\rm{ref}}\right)^{-1}$. Thus, denoting the second matrix in Eq.~\ref{Eq.23} by $\boldsymbol\kappa$ , we obtain
\begin{align}\label{eq_for_kappa}
\boldsymbol\kappa=\mathbf{B_1^k}\left(\mathbf{C_{\rm{ref}}^k}\right)^{-1}.
\end{align}

In summary, the entire RIXS calculation can be efficiently executed by initially calculating and retaining the reference matrices $\mathbf{B_1^k}$ , $\mathbf{B_{\rm{ref}}^k}$ , $\mathbf{C_{\rm{ref}}^k}$, from which we can obtain $D_1^{k,\alpha,\beta}$ ($D_2^{k,\alpha,\beta}$) using Eq.~\ref{eq17} and~\ref{eq_for_gamma} (Eq.~\ref{Eq.21} and~\ref{eq_for_kappa}). $A^{a}_{\rm{GS}\rightarrow X_k}$ is found with the help of Eq.~\ref{eq7}. With these ingredients, we can find the RIXS cross-section by plugging into Eq.~\ref{eq1} the revised form of Eq.~\ref{eq1a}, which now reads 

\begin{align}
&{}T_{\omega_{\rm{in}},F} =\\
\nonumber
&{} \sum_k \frac{\left[\sum_{\alpha,\beta}  \left(C^F_{\alpha,\beta}\right)^* \left(D_1^{k,\alpha,\beta} - o^*_\beta D_2^{k,\alpha}\right)\right] \cdot A^{a}_{\rm{GS}\rightarrow X_k}}{\omega_{\rm{in}}-(E_k - E_{\rm{GS}})+i\Gamma},
\end{align}
where, as before, we have omitted the polarization-dependence for simplicity.


\section{RESULTS}

\begin{figure}
\centering
\includegraphics[width=0.49\textwidth]{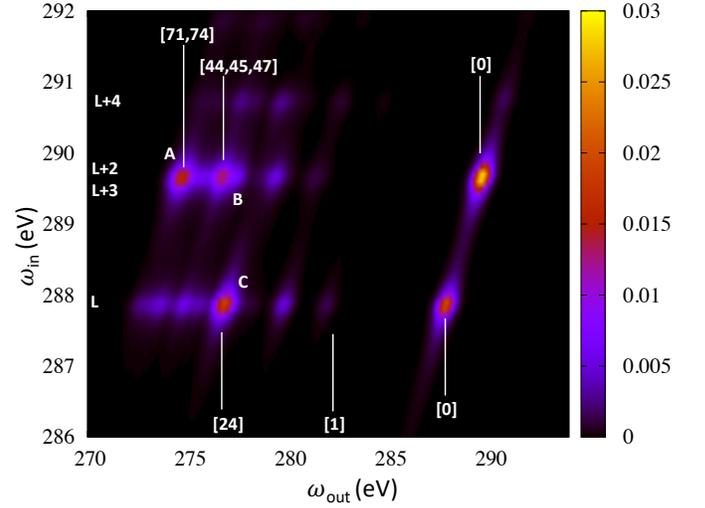}
\caption {Carbon K-edge RIXS spectrum of methanol simulated within the CleaRIXS formalism employing the PBE functional. The horizontally aligned bright features at $\sim$ $\omega_{\rm{in}} = $ 287.8 eV, 289.6 eV and 290.7 eV contain contributions predominantly from intermediate states where the core electron has been excited to the LUMO (L) orbital , the near-degenerate LUMO+2 (L+2) and LUMO+3 (L+3) orbitals and the LUMO+4 (L+4) orbital, respectively. The indices of the final states contributing heavily to certain high-intensity features are mentioned in square brackets, where [0] denotes the ground state and [$n$] denotes the $n$-th excited state of the molecule. Note that, as expected, the intense features along a given vertical line do not correspond to de-excitation to the same final state.}\label{Fig_1}
\end{figure}

As a demonstrative example, in Fig.~\ref{Fig_1}, we show the simulated C K-edge RIXS map of an isolated $\mathrm{CH_3OH}$ (methanol) molecule with respect to $\omega_{\rm{out}}$ and $\omega_{\rm{in}}$ , respectively. 
The necessary transition-matrix $(\{o_m\})$ and overlap $(\{\xi_{i,j}\})$ terms are evaluated from the relevant KS orbitals, which, are calculated with the {\sc Quantum} ESPRESSO software-package employing the Perdew-Burke-Ernzerhof (PBE) functional. The energies $\{E_F\}$ and coefficients $\{C^F_{\alpha,\beta}\}$ for the final valence-excited state (see Eq.~\ref{f_expansion}), which can, in principle be calculated by applying any  response-based approach (like LR-TDDFT/BSE) on the ground state SD, are obtained using the ($\mathrm{G_0W_0}$ + adiabatic BSE) formalism, implemented within the BerkeleyGW package~\cite{DESLIPPE20121269}. 

\subsection{Feature Analysis}

We can divide Fig.~\ref{Fig_1} into spectral features along various diagonal and horizontal lines: 

(1) Those along the diagonal line $\omega_{\rm{in}} = \omega_{\rm{out}}$ correspond to elastic scattering ($\ket{F}=\ket{\rm{GS}}$). 

(2) Each point on a given horizontal line can roughly be associated with a specific core-excited state. In principle, all intermediate core-excited states $\{\ket{X_k}\}$ are included in the definition of $T_{\omega_{\rm{in}},F}$ (Eq.~\ref{eq1a}), however, the denominator makes sure that, for a given $\omega_{\rm{in}}$, only those intermediate states for which $\omega_{\rm{in}} \approx (E_k - E_{\rm{GS}})$, within a small energy window controlled by $\Gamma_k$ can have appreciable contribution. For example, the intermediate state with an excited electron in the LUMO (LUMO+4) orbital is primarily responsible for the intense features on the horizontal line at $\sim$ $\omega_{\rm{in}}=287.8$ (290.7) eV, while the horizontally arranged intense features at $\sim$ $\omega_{\rm{in}}=289.6$ eV are mostly due to intermediate states having an electron in the near-degenerate orbitals (LUMO+2) and (LUMO+3).

(3) In Eq.~\ref{eq1}, the energy conservation term $\delta(\omega_{\rm{out}}-\omega_{\rm{in}} -[E_F -E_{\rm{GS}}])$ arranges specific final state features along diagonals displaced by $E_F -E_{\rm{GS}}$ to the left of the elastic line. In practical calculations, this delta-function is replaced by a Gaussian with a half-width $\sigma$ which blurs together final states within this small energy window.

(4) By this logic, vertically aligned features in the RIXS plane (Fig.~\ref{Fig_1}), i.e., for fixed $\omega_{\rm{out}}$ typically correspond to completely different final states, at least for purely electronic, near-edge absorption (input) energies. (We provide more discussion on vibrational coupling and the transition to non-resonant XES below.)
For example, the 24-th valence-excited state, in which the most prominent contribution is a single-electron transition from the (HOMO-3) to the LUMO orbital of the ground state, is predominantly responsible for the intense feature C (Fig.~\ref{Fig_1}) at $(\omega_{\rm{in}} = 287.8\textrm{ eV}$ ,  $\omega_{\rm{out}} = 276.8\textrm{ eV})$. On the other hand, the bright feature B at $(\omega_{\rm{in}} = 289.6\textrm{ eV}$ , $\omega_{\rm{out}} = 276.8\textrm{ eV})$ on the same vertical line originates mostly from de-excitations to the the 44-th (main contribution : HOMO-2 $\rightarrow$ LUMO+3) , the 45-th (main contributions : HOMO-2 $\rightarrow$ LUMO+2 , HOMO-2 $\rightarrow$ LUMO+3 and HOMO-2 $\rightarrow$ LUMO+4) and the 47-th  (main contributions : HOMO-3 $\rightarrow$ LUMO+4) final valence-excited states.

\begin{figure}
\centering
\includegraphics[width=0.49\textwidth]{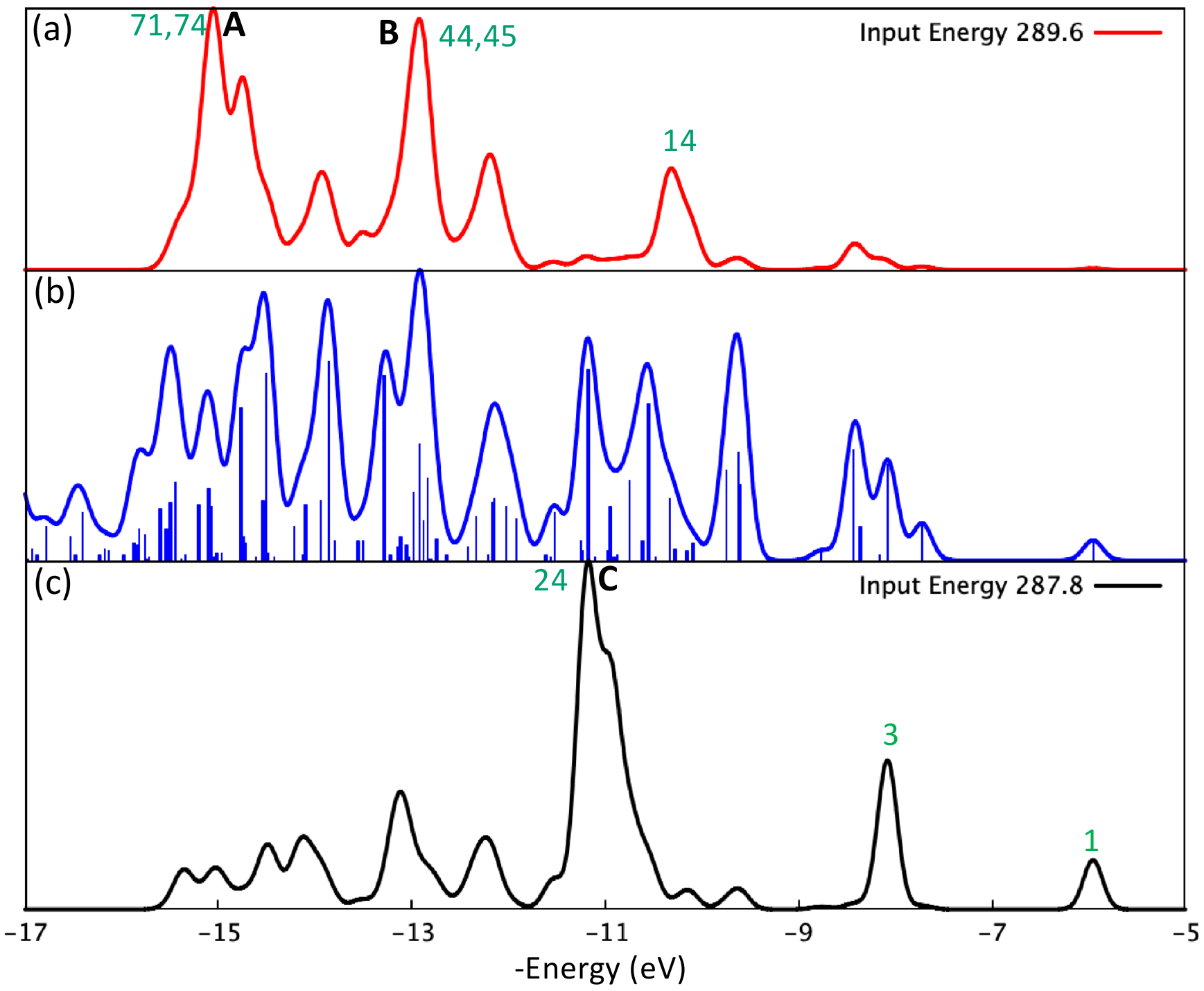}
\caption{Panels (a) and (c) show the CleaRIXS emission spectra of intensity (y-axis) vs $\omega_{\rm{out}} - \omega_{\rm{in}}$ (x-axis) corresponding to $\omega_{\rm{in}} = 289.6$ eV and $\omega_{\rm{in}} = 287.8$ eV, respectively. The numbers beside the peaks represent the indices of the final excited states with appreciable contribution. Certain high-intensity peaks are labeled as A , B and C. Panel (b) shows the valence excitation spectrum of the ground-state system with the horizontal axis corresponding to the negative of the excitation energy. The columns in panel (b) show the oscillator strengths (in arbitrary units) of the excitations. Note that a vertical line through all three panels can be associated with a given valence-excited final state.}\label{Fig_2}
\end{figure}

\begin{figure}
\centering
\includegraphics[width=0.46\textwidth]{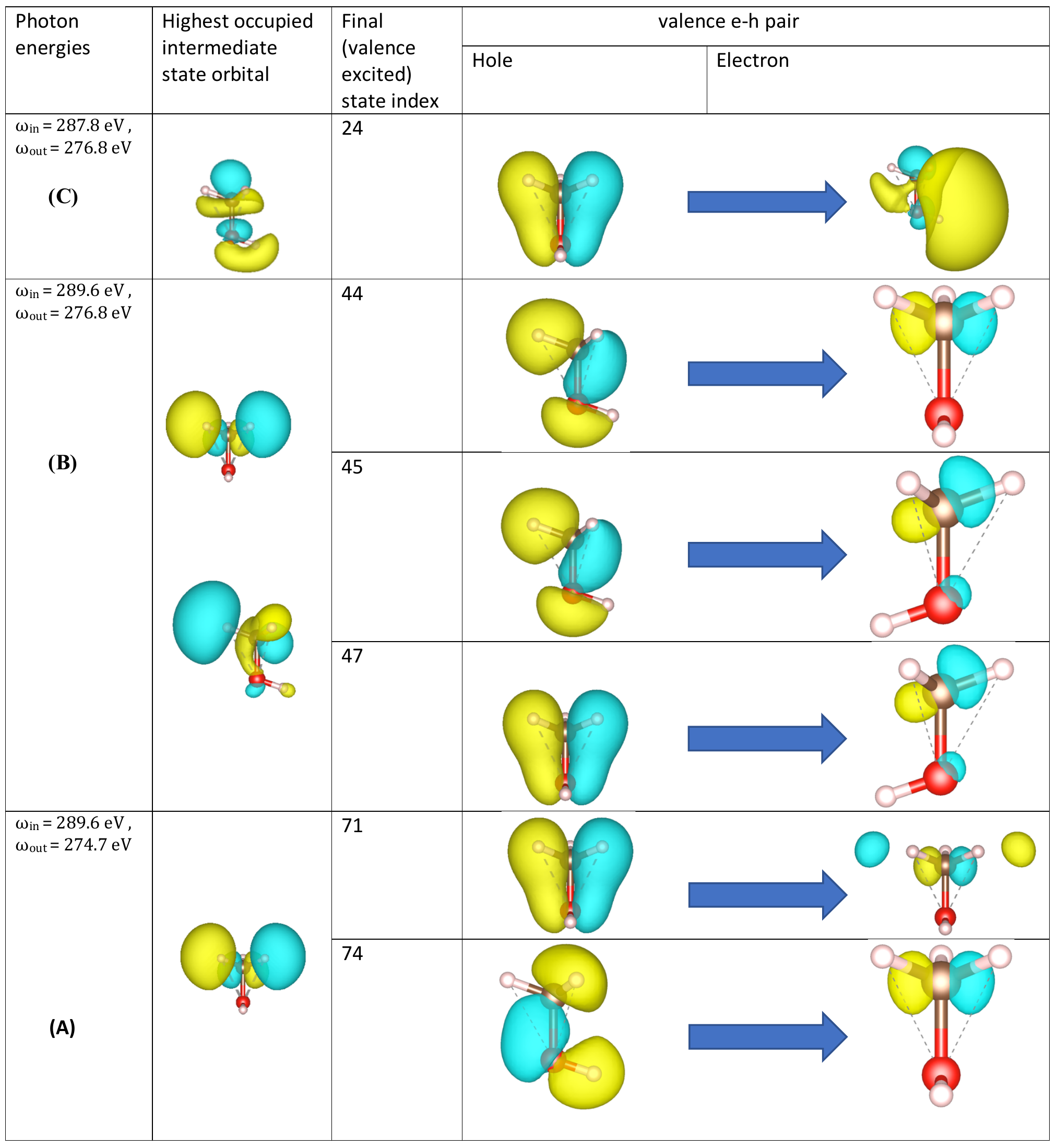}
\caption {Table showing analysis of certain high-intensity features (labeled in accordance with Fig.~\ref{Fig_1},~\ref{Fig_2}) of the CleaRIXS map. The features are specified with the incident and emitted photon frequency in column 1. Column 2 shows the orbital of the excited electron in the intermediate state. The index of the final valence-excited state is presented in column 3, with index=0 corresponding to the ground state. The last column shows the dominant term in the linear-combination of electronic transitions constituting the final state, with the arrow pointing in the direction of the promotion of electron from an occupied to an unoccupied ground state orbital (which, in the final state, are a hole and an excited electron, respectively.) }\label{FigTable}
\end{figure}

The interrelation between the RIXS spectral features and the various valence-excited final states can be better appreciated from Fig.~\ref{Fig_2}. Panels (a) and (c) show the emission intensity plotted as a function of inelastic energy loss, $\omega_{\rm{out}}-\omega_{\rm{in}}$,  or its negative in this specific case, with loss increasing to the left, for $\omega_{\rm{in}}=$ 289.6 eV and 287.8 eV, respectively. From Eq.~\ref{eq1}, we can see that, for such plots, regions with the same energy loss must represent the same or degenerate final states\footnote{replacing the delta function with a Gaussian will include contributions from other final states within a small energy-window.}, with energies referenced to the ground state, $(E_F - E_{\rm{GS}})$. Panel (b) shows, as a function of the negative of the excitation energy, the oscillator-strengths of the valence-excited states for transitions from $\ket{\rm{GS}}$. It is important to note that these oscillator strengths do not resemble the intensities plotted in panels (a) or (c). This is expected since the emission intensity is related to the oscillator strength of a different transition, namely that from $\ket{X_k}$ to $\ket{F}$. It is in this way that states that are inaccessible (dark) in optical excitations can often be probed using RIXS. 

\subsection{Orbital Analysis}

In order to provide a visual representation of the excitation--de-excitation processes accessible within the CleaRIXS framework, in Fig.~\ref{FigTable}, we show the isovalue plots of the orbitals dominating the electronic transitions for certain intense features of the RIXS map. In particular, as shown in the first column, we analyze the RIXS features present at C ($\omega_{\rm{in}}=$ 287.8 eV , $\omega_{\rm{out}}=$ 276.8 eV) , B ($\omega_{\rm{in}}=$ 289.6 eV , $\omega_{\rm{out}}=$ 276.8 eV) and A ($\omega_{\rm{in}}=$ 289.6 eV , $\omega_{\rm{out}}=$ 274.7 eV) and plot the corresponding highest occupied orbitals in the core-excited state $\ket{X_k}$ in column 2. Note that multiple final states within a narrow energy-window can contribute to the same RIXS feature. Column 3 denotes the indices, with reference to the ground-state, of such valence excited-states with appreciable contributions. Each valence excited-state, in turn, is expressible as a linear combination of single-electron transitions from an occupied to an unoccupied ground-state orbital. The final column provides a visual representation of the single-electron transition with the largest contribution such that the arrow points from the occupied GS orbital (i.e., the final-state hole) to the unoccupied GS orbital (i.e., the excited electron in the final state).

\begin{figure}
\centering
\includegraphics[width=0.49\textwidth]{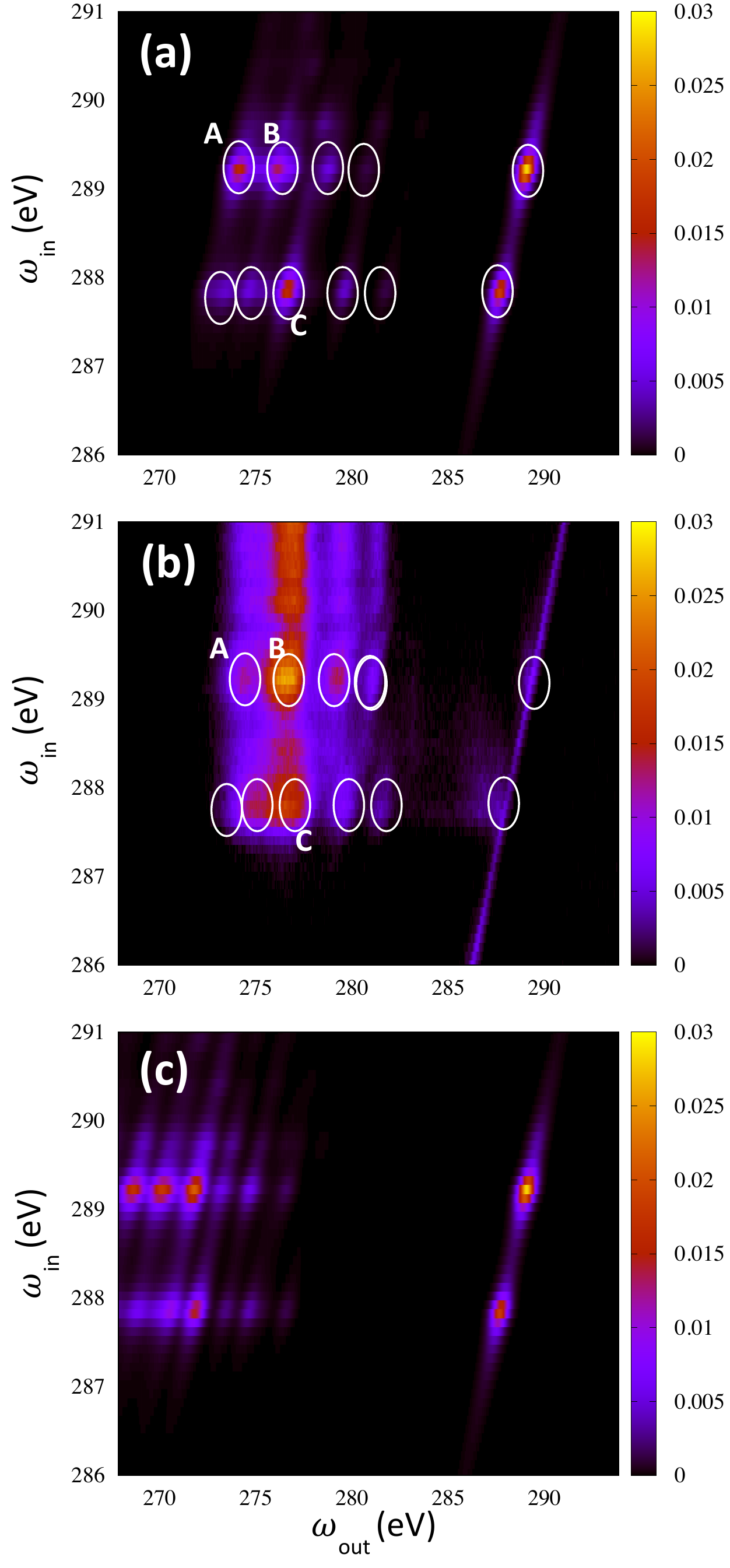}
\caption{Panel (a) shows the simulated CleaRIXS spectrum. Eigenvalues of the BSE are used as the final (valence-excited) state energies while the intermediate (core-excited) state energies are obtained from the relevant $\mathrm{G_0W_0}$ eigenvalues. Panel (b) shows the experimental spectrum while the simulated spectrum calculated neglecting the valence e-h interaction (such that all energies are obtained from the relevant $\mathrm{G_0W_0}$ eigenvalues) is presented in panel (c). White rings are used to highlight some bright features in panels (a) and (b). The relative position of the rings are kept unaltered between the two panels, corroborating the agreement in energy positions.}\label{fig_TheoryExperiment}
\end{figure}

\subsection{Role of Electron-Electron Interactions}
In order to explore the effects of valence e-h interactions in the final state, in Fig.~\ref{fig_TheoryExperiment}(a) and (c), we present the simulated RIXS spectra obtained respectively by incorporating and ignoring the valence e-h interaction in the final state. In other words, the final states in panel (a) are obtained with accurate BSE calculations as described earlier. In contrast, each final state in panel (c) is obtained by creating a hole in an occupied frozen GS orbital and an electron in an empty frozen GS orbital (with the energy eigenvalue replaced by the $\mathrm{G_0W_0}$ counterpart). Note, that the large separation between the elastic and the inelastic region is consistent with the large $\mathrm{G_0W_0}$ quasiparticle (QP) gap of methanol~\cite{doi:10.1021/acs.jctc.5b00453}. 
To ensure higher accuracy, in both of the panels (a) and (c), the KS eigenvalues of the intermediate state are replaced by the $\mathrm{G_0W_0}$ counterparts~\cite{PhysRevB.34.5390}. Note that Fig.~\ref{Fig_1} differs from Fig.~\ref{fig_TheoryExperiment}(a) in that the former does not include the $\mathrm{G_0W_0}$ corrections to the intermediate-state KS eigenvalues.

The experimental RIXS spectrum, reported in ref.~\cite{doi:10.1021/acs.jpca.6b02636}, is shown in panel (b). Note that inclusion of the valence e-h interactions significantly improves the agreement of the energy of the intense features of the simulated results with experiment. For example, in panel (c), for $\omega_{\rm{in}}=$ 287.8 eV, the feature corresponding to the lowest-energy final state, which essentially corresponds to promotion of an electron from the ground state HOMO to the LUMO orbital, is found at $\omega_{\rm{out}}=276.7$ eV. In contrast, as shown in panel (a), owing to the significant valence e-h binding, the lowest energy final excited state is found, in agreement with the experimental plot (b), at  $\omega_{\rm{out}}=281.7$ eV for the same $\omega_{\rm{in}}$. 
In particular, at least in the low $\omega_{\rm{in}}$ region, the input and the output energy values corresponding to the bright features show excellent agreement between panels (a) and (b). The relative difference seen in the high $\omega_{\rm{in}}$ region is discussed in the following sections.

\begin{figure}
\centering
\includegraphics[width=0.49\textwidth]{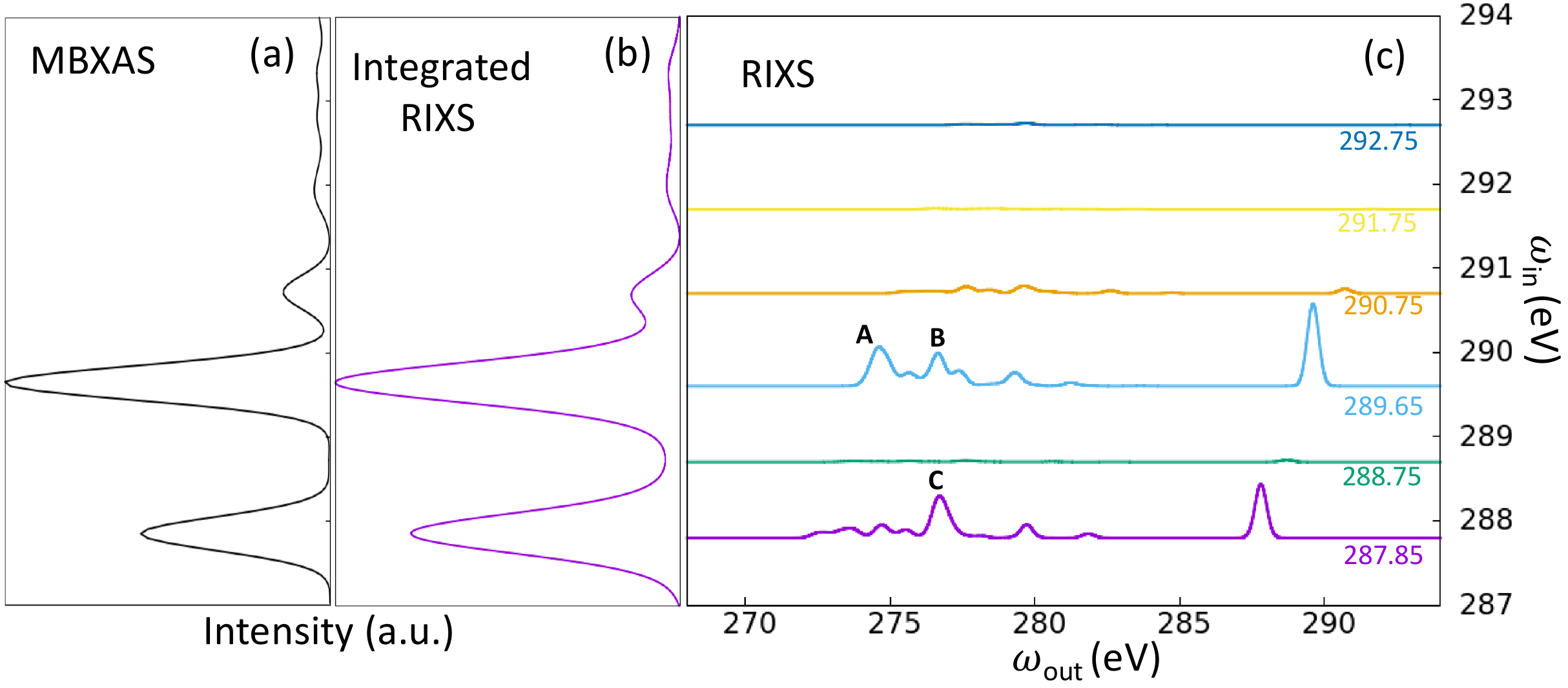}
\caption{The C K-edge absorption spectrum of methanol obtained using the MBXAS formalism is shown in panel(a). Panel (b) shows, as a function of $\omega_{\rm{in}}$, the simulated emission intensity integrated over $\omega_{\rm{out}}$. Panel (c) shows the simulated RIXS spectra corresponding to different values of $\omega_{\rm{in}}$.}\label{fig_XAS_RIXS}
\end{figure}

\subsection{Connection with XAS}
An approximate relation between the RIXS spectrum and the conventional XAS spectrum can be established by noting that for a given input energy, the spectrum of observed emission must originate from the accessible intermediate states. So, the integrated emission intensity corresponding to any input energy $\omega_{\rm{in}}$ should be proportional to the associated XAS intensity. Under such circumstances, for any given $\omega_{\rm{in}}$, the approximate XAS intensity can be obtained by integrating the RIXS spectrum over $\omega_{\rm{out}}$. As seen in Fig.~\ref{fig_XAS_RIXS}, the C K-edge XAS of methanol (panel a) simulated using the many-body X-ray absorption spectroscopy (MBXAS) method~\cite{PhysRevLett.118.096402,PhysRevB.97.205127} is in very good agreement with the CleaRIXS spectrum integrated over  $\omega_{\rm{out}}$ (panel b). The emission spectra corresponding to some selected values of $\omega_{\rm{in}}$ are shown in panel (c). Note that unlike panels (a) and (c) of Fig.~\ref{Fig_2}, in which the x-axis stands for $\omega_{\rm{out}} - \omega_{\rm{in}}$ , in Fig.~\ref{fig_XAS_RIXS}(c), the x-axis is simply $\omega_{\rm{out}}$. Consequently, the peaks B and C, which have different energy loss in Fig.~\ref{Fig_2}, coincide here. Since the XAS intensity is proportional to $\omega_{\rm{in}} \sum_k| \braket{X_k|\hat{O}|\rm{GS}}|^2 \delta(\omega_{\rm{in}}-(E_k-E_{\rm{GS}}))$, a low XAS intensity corresponding to some input frequency $\omega'_{\rm{in}}$ would typically imply low value of $| \braket{X_k|\hat{O}|\rm{GS}}|$ for any intermediate state for which $E_k-E_{\rm{GS}} \approx \omega'_{\rm{in}}$. Therefore, from Eq.~\ref{eq1a}, one can conclude that, for such an input frequency $\omega'_{\rm{in}}$, the RIXS spectrum will not contain any intense emission peak. Such a trend is easily observed in Fig.~\ref{fig_XAS_RIXS}.

\subsection{Additional Features in the Experimental Spectra}

A comparison between the simulated (Fig.~\ref{fig_XAS_RIXS}(a)) and experimental (Fig.~4 of  ref.~\cite{doi:10.1021/jp0219045}) XAS spectra reveals that the experimental counterpart contains several additional features of appreciable intensity, especially beyond the strong peak at $\sim289.5$  eV, which are not present in the simulated spectrum. Such difference in the absorption spectra, which are also reflected in the RIXS results (especially in the high $\omega_{\rm{in}}$ region), can be attributed to the following effects which are beyond the scope of this paper:
\begin{enumerate}
    \item The high-energy unoccupied levels of the intermediate state can have appreciable contributions from the $p$-orbitals of higher principal quantum number, which are not included in the projector-augmented wave (PAW) based treatment~\cite{PhysRevB.66.195107} used for computing the single-particle transition dipole moments. Such underestimation of the absorption intensity in the high-energy region has been reported in several studies~\cite{C4CP05316H, doi:10.1021/jacs.5b05854, https://doi.org/10.1002/sia.5895, doi:10.1063/1.4856835, doi:10.1021/jp806823c} employing the PAW-based approach.
    \item In addition to the purely electronic transitions considered in the simulations, the incident photon can excite combined (electronic+vibrational) states, giving rise to additional peaks. For example, between the absorption peaks at $\sim287.9$ eV and $\sim289.5$ eV, the experimental XAS spectrum shows multiple features  which have been associated mostly with combined electronic and vibrational effects. Similar effects are said to be responsible for several of the intense features beyond the sharp peak at $\sim289.5$ eV.
\end{enumerate}

In the context of RIXS, if, besides the promotion of the core-electron, additional vibrational excitations are created in the intermediate state, then the final valence-excited state is also likely to contain vibrational excitations~\cite{PhysRevLett.110.265502}. If the energy associated with the vibrational excitation is similar in the intermediate and the final state, then, in the corresponding region in the RIXS map, an increment in $\omega_{\rm{in}}$ (needed to create the additional vibronic excitations) will not result in any appreciable change in $\omega_{\rm{out}}$, leading to vertically aligned features.
%

\subsection{Connection with Non-resonant XES}
As the incident energy extends above the absorption edge, the non-resonant X-ray emission spectrum emerges revealing decay channels for the occupied valence subspace while the excited electron is emitted as an unbound final state. Formally, we can see this connection in the Kramers-Heisenberg expression (Eqs.~\ref{eq1} and~\ref{eq1a}). We partition the sums over states to make a distinction between bound and unbound excited states, using estimates of the first ionization potentials of the valence and core electron subspaces (for a specific core-excited atom). The unbound states are written as products of $N-1$ electron stationary states with the emitted/unbound electron represented as one wave with a well-defined energy from a continuous basis defined by the symmetry of the system (whether a finite molecule or cluster, a semi-infinite surface or a periodic condensed phase).

In general, for a RIXS map, if IE is the first (or valence) ionization-energy of the system, then any feature to the right hand side of the line $\omega_{\rm{in}} - \omega_{\rm{out}} = \textrm{IE}$ can only arise from bound final states. If $-\varepsilon_\alpha$ is the QP energy of a valence occupied orbital $\ket{\alpha}$, then any final state resulting solely from the ionization of this orbital can, in principle, contribute vertical features at points satisfying the condition $(\omega_{\rm{in}} - \omega_{\rm{out}}) > \varepsilon_\alpha$. This, in fact, is true for points to the left of a straight line given by $(\omega_{\rm{in}} - \omega_{\rm{out}}) = \varepsilon_\alpha$, which is parallel to the elastic line.

\begin{figure}
\centering
\includegraphics[width=0.45\textwidth]{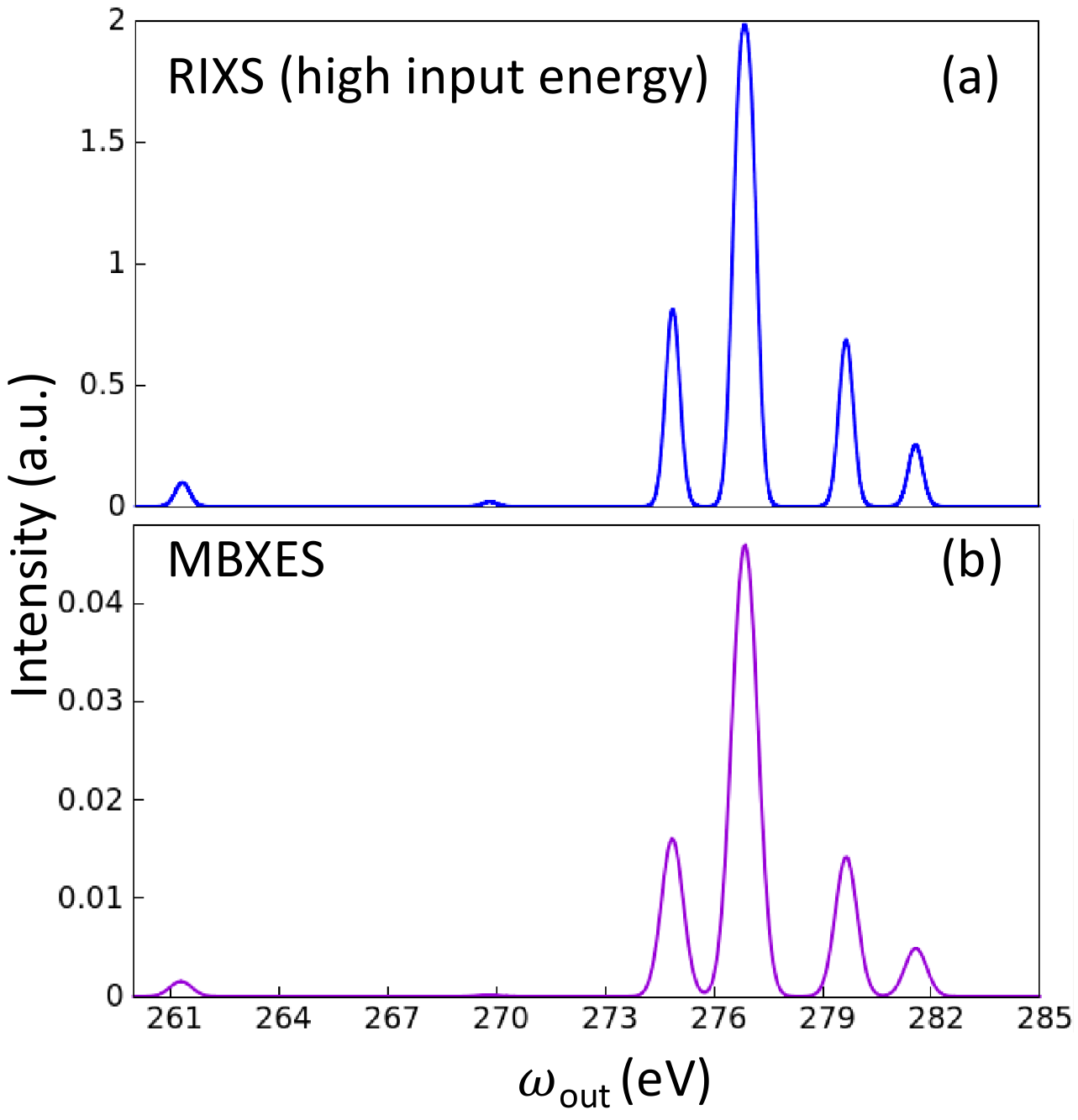}
\caption{Panel (a) shows the simulated emission spectrum, calculated within the KS-RIXS framework, corresponding to $\omega_{\rm{in}} = 307.3$ eV obtained in conjunction with the condition that only those intermediate and final states in which the highest conduction orbital is occupied, are allowed to contribute. Panel (b) shows the non-resonant X-ray emission spectrum obtained with the MBXES method.}\label{Fig_XES_RIXS}
\end{figure}

If $\omega_{\rm{in}} \geq$ core-ionization energy (or ionization potential, IP), any dominant intermediate state, with energy $E_{\rm{FCH}}+\varepsilon_{k'}$, can be written as $\ket{\rm{FCH}}\otimes\ket{k'}$, where $\varepsilon_{k'}$ is the energy of the unbound electron in orbital $\ket{k'}$ and $E_{\rm{FCH}}$ is the energy of the lowest core-ionized (full core-hole) state $\ket{\rm{FCH}}$. For sufficiently high $\omega_{\rm{in}}$, $\ket{k'}$ will be a scattering state with high-enough energy such that, the emission amplitude will be negligible unless the final state contains an unbound electron and can be written as $\ket{F}=\ket{F'}\otimes\ket{f'}$, where $\ket{f'}$ ($\varepsilon_{f'}$) is the state (energy) of the unbound electron and $\ket{F'}$ ($\varepsilon_{F'}$) is the state (energy) of the rest of the system, comprising bound electrons. Under such circumstances, the RIXS cross-section can be written as
\begin{align*}\label{EQ_RIXStoXES1}
    \sigma_{p,q}(\omega_{\rm{in}},\omega_{\rm{out}}) &{}= \sum_{F'} \left\rvert\braket{F'|O_p^\dagger|\rm{FCH}}\right\rvert^2\\ &{}\sum_{f'} \delta(\omega_{\rm{in}}+E_{\rm{GS}}-\omega_{\rm{out}}-E_{F'}-\varepsilon_{f'})\\
    &{}\left\rvert\sum_{k'}\frac{\braket{f'|k'}\bra{k'}\otimes\braket{\rm{FCH}|O_q|\rm{GS}}}{\omega_{\rm{in}}-E_{\rm{FCH}}-\varepsilon_{k'}+E_{\rm{GS}}+i\Gamma}\right\rvert^2.\numberthis
\end{align*}
Recognizing that in the non-resonant limit the initially emitted electron $\ket{k'}$ and the final unbound electron $\ket{f'}$ must be the same electron, we can simplify this expression significantly to
\begin{align}\label{EQ_RIXStoXES}
    \sigma_{p,q}(\omega_{\rm{in}},\omega_{\rm{out}}) &{} \propto \sum_{F'} \left\rvert\braket{F'|O_p^\dagger|\rm{FCH}}\right\rvert^2 \delta(\omega_{\rm{out}}+E_{F'}-E_{\rm{FCH}}),
\end{align}
with $\ket{F'}$ and $\ket{\rm{FCH}}$ approximated by the corresponding SDs. Eq.~\ref{EQ_RIXStoXES} is the expression used for simulating non-resonant XES (NXES) in the many-body X-ray emission spectroscopy (MBXES) method~\cite{roychoudhury2021changes}.

Fig.~\ref{Fig_XES_RIXS}(a) shows the emission spectrum for $\omega_{\rm{in}}$ = 307.3 eV obtained using Eq.~\ref{EQ_RIXStoXES1} where both $k'$ and $f'$ are restricted to the highest available orbital. Panel (b) shows the NXES spectrum obtained with MBXES.

\section{Conclusion}
In conclusion, in this paper we introduce CleaRIXS, an accurate computational framework for simulating the RIXS spectrum by treating different levels of e-h interactions using different techniques. In the intermediate state, the attraction between the core hole and the excited electron is treated within the SCF approach employing a non-aufbau constrained occupation. On the other hand, the valence e-h interaction in the final state is incorporated using the linear-response formalism. Even though the CleaRIXS method, in its current state, considers only electronic excitations in the final state, extensions can be developed in a straightforward way to take additional excitations (e.g., ionic vibrations) into account. We show that, in addition to offering a fast and reliable simulation of the RIXS process, the aforementioned partitioning of the calculation enables us to easily map the RIXS features to the corresponding electronic excitations and de-excitations. Using the C K-edge spectrum of methanol as an example, we compare the simulated spectrum with the experimental one and demonstrate the requirement of the linear-response treatment for the final state. We qualitatively rationalize the presence of additional vertical features in the experimental map as results of vibrational excitations. Finally, we show, with theoretical considerations and computed examples, that, as limiting cases, the spectrum obtained with CleaRIXS matches with the XAS and NXES spectra obtained using recently developed determinant-based techniques. As a continuation of our development, in near future we plan to extend CleaRIXS to simulate RIXS in periodic systems and to incorporate ancillary functionalities to take into account additional excitations (e.g. vibrations, charge-transfer excitations, d-d excitations) in the final core-filled state.

\section{Acknowledgement}

This work was performed at the Molecular Foundry, LBNL and was supported by the Office of Science, Office of Basic Energy Sciences, of the U.S. Department of Energy under Contract No. DE-AC02-05CH11231. Computational works were carried out using supercomputing resources of the National Energy Research Scientific Computing Center (NERSC) and the TMF clusters managed by the High Performance Computing Services Group, at LBNL. S.R. acknowledges the help of Zhenglu Li (UC Berkeley) with queries related to  BerkeleyGW and Daniele Varsano (CNR, Modena) for helpful discussions. The authors are thankful to Lothar Weinhardt (Karlsruhe Institute of Technology) for providing the data for the experimental RIXS spectrum.

\bibliography{biblio.bib}
\end{document}